\documentclass[12pt,preprint]{aastex}

  \def \lta {\mathrel{\vcenter
     {\hbox{$<$}\nointerlineskip\hbox{$\sim$}}}}
\def \gta {\mathrel{\vcenter
     {\hbox{$>$}\nointerlineskip\hbox{$\sim$}}}}
\def \m{\ifmmode M_\odot\else M$_\odot$\fi}
\def \l{\ifmmode L_\odot\else L$_\odot$\fi}
\def \r{\ifmmode R_\odot\else R$_\odot$\fi}

\def\gcm3{g~cm$^{-3}$}
\def\g-s{g~s$^{-1}$}
\def\cm3s{cm$^3$~s$^{-1}$}

\def\erg-s{erg~s$^{-1}$}

\def\beq{\begin{equation}}
\def\eeq{\end{equation}}
\def\gr{$\gamma$-ray}
\def\grs{$\gamma$-rays}

\def\grbs{$\gamma$-ray bursts}
\def\0{\parindent=0cm}
\def\5{\parindent=.5cm}
\def\7{\parindent=.7cm}

\begin{document}

\title{ASTROPHYSICAL AND ASTROBIOLOGICAL IMPLICATIONS OF GAMMA-RAY BURST
PROPERTIES}

\author{John Scalo and J. Craig Wheeler}

\affil{The University of Texas at Austin}

\begin{abstract}

\setcounter{page}{1}

Combining results from Schmidt (1999) for the local cosmic rate
and mean peak luminosity of \grbs\ (GRBs) with results on the
history of the cosmic star formation rate, we provide estimates
for the {\it local} GRB rate per unit blue luminosity in galaxies.
For a moderate increase in SFR with redshift, we find a GRB rate per
unit B luminosity of ${\rm 2.4\times10^{-17} h_{70}^2~\l^{-1}~yr^{-1}}$.
The corresponding mean \gr\ luminosity density in the 
Milky Way is $1.6\times10^{29}$ \erg-s~pc$^{-2}$ and the total
rate is $5.5\times10^{-7} h_{70}^2$~yr$^{-1}$.  
These values are used to examine a number of phenomena with the
following conclusions: 1) The ratio of supernova rate to isotropic
equivalent GRB rate
is large: ($\gta$ 6000 SN~Ibc per GRB, $\gta$ 30,000 SN~II per GRB).
With no correction for collimation
it is difficult to maintain that more than a small
fraction of neutron star or black hole-forming events produce GRBs.
GRBs could arise in a large fraction of black hole or magnetar-forming
events only with collimation in the range $\Delta \Omega/4\pi 
\sim 0.01 - 0.001$ and a steep enough slope of the IMF;
2) Without substantial collimation, the GRB rate is small; with
collimation, the energy input is small.  The net effect is
that it is impossible to use these events to account for the majority
of large HI holes observed in our own and other galaxies; 3) Modeling the
GRB events in the Milky Way as a spatial Poisson process and allowing for
modest enhancement in the star formation rate due to birth in a spiral arm,
we find that the probability that the solar system was
exposed to a  fluence large enough to melt the chondrules during the
first 10$^7$ yr  of solar system history is negligibly small,
independent of collimation effects.  This is
especially true considering that there is strong evidence that the
chondrules were melted more than once; 4) We calculate the
probability that surfaces of planets and satellites have been subjected to
irradiation from GRBs at fluence levels exceeding those required for
DNA alterations during a given  period of time. 
Downscattering to energies at which photoelectric absorption occurs
results in a transmission factor for ionizing radiation that is an 
exponential function of the atmospheric column density.  Even for
very opaque atmospheres, a significant fraction of the GRB energy
is transmitted as UV lines due to excitation by secondary electrons.
For eukaryotic-like organisms in thin atmospheres (e.g. contemporary
Mars), or for UV line exposure in thick atmospheres (e.g. Earth)
biologically significant events occur at a rate of $\sim$ 100 -- 500
Gyr$^{-1}$.  The direct contribution of these ``jolts" to 
mutational evolution may, however, be negligible because of the
short duration of the GRBs.  Evolutionary effects due to 
partial sterilizations and to longer-lived
disruptions of atmospheric chemistry should be more important.

\end{abstract}


\section{INTRODUCTION}

The discovery of large redshifts for some \grbs\
(GRBs), through redshifts of afterglow lines or association
with galaxies, showed that the intrinsic \gr\ luminosities
must be very large.  A succinct summary is given by van
Paradijs (1999).  These large photon energies, and the implied
large associated kinetic energies, have led several workers to
suggest that GRBs might be responsible for a number of observed
astrophysical phenomena.  These include the production of
numerous large HI holes observed in our own and other galaxies
(Efremov, Elmegreen \& Hodge 1998; Loeb \&
Perna 1998) and the melting of dust grains resulting in the
formation of chondrules in the early solar system (McBreen \&
Hanlon 1999).  These issues depend sensitively on the assumed
luminosities or kinetic energies of GRBs and especially on their
rates.  Estimates of these quantities are possible because of
the mounting evidence that GRBs are associated with massive
star precursors.  This is
suggested by the presence of GRBs near the centers of galaxies
with active star formation (Hogg \& Fruchter 1999) and
especially the light curve signatures detected in
afterglow observations of GRB 980326 (Bloom et al.~1999) and
GRB 970228 (Reichart 1999; Galama et al.~2000) that
are consistent with supernovae (SN), as well as the
coincidence between GRB 980425 and SN 1998bw (Galama et al.~1998).
For this reason it is believed that GRBs
(at least those of long duration) track the star
formation rate (SFR) in galaxies.  Some assumptions about the
average cosmic SFR history of the Universe as a function of
redshift then allow a comparison of models with number-flux
counts and other statistical constraints in order to derive
average peak GRB luminosities and rates.

Earlier work used GRB rates derived by assuming that all GRBs
have the same luminosity (the standard candle assumption),  as
in Wijers et al. (1998).  In addition, it was originally thought that the
cosmic SFR
is a rapidly increasing function of redshift, as estimated primarily from UV
luminosities (see Lilly et al.~1996, Madau et al.~1996,
Connolly et al.~1997).  Both assumptions are very uncertain,
and further work has shown how to improve upon them.  First,
the measurement of redshifts has shown that GRBs are certainly
not standard candles.  Schmidt (1999) has calculated the mean peak
luminosities and cosmic GRB rates per unit volume using a variety
of assumed peak luminosity functions, as well as two choices of
redshift evolution and cosmological parameter, $q_0$, thus
eliminating the need for the standard candle assumption.
Second, a careful analysis of the derivation of the UV
luminosity density using deep spectroscopic observations by
Cowie, Sangaila \& Barger (1999) has considerably reduced the increase of
the cosmic SFR with redshift out to z$\sim$1 compared to earlier
estimates.  A similar conclusion was reached in the
spectroscopic study of compact galaxies with z$<$1.4 by Guzman
et al. (1997), using [OII] equivalent widths to estimate the SFR
for z$>$0.7 and z$<$0.7.

In the present paper, we use these two results to estimate more
reliable values of the GRB mean peak luminosity and the rate
per unit host galaxy blue luminosity and find that they are fairly
well-constrained (\S2).  These quantities are then used to re-examine
a number of questions.  We compare the GRB rates with galactic
SN rates in order to constrain the fraction of SN that yield
GRBs (\S3); only with generous allowance for collimation, and a
favorable IMF slope, is it possible to maintain that more than
a small fraction of neutron star or black hole-forming events
produced GRBs (\S4).  We then consider the likelihood that GRBs are
responsible for most of the HI holes in galaxies and for the
melting of chondrules in the early solar nebula (\S5).  We reach
negative conclusions, not so much because of the revised rates
and luminosities, but because of empirical constraints not
considered in previous work.  We do find that GRBs are capable
of supplying an intermittent  terrestrial surface fluence in
excess of that required for direct biological effects (DNA
alterations) with a mean time interval of about 10$^7$ yr
for eukaryotes (\S6). 
We discuss the possible consequences for biological evolution.

\section{GRB GALACTIC RATES AND MEAN PEAK LUMINOSITIES}

        For the applications considered below, we would like to know
separately the rate of occurrence of GRBs in single galaxies of given
characteristics, in particular luminosity, and the mean peak luminosity
of GRBs.  For some applications we require these two quantities
separately, in others they occur as a product.  Because there are too few GRBs
with known redshift to directly construct a luminosity function or
estimate a rate, an indirect method relying only on observed fluxes
is required.  

Two methods for obtaining GRB rates and peak luminosities
that are formally equivalent are to
use the distribution of $V/V_{max}$, or, alternatively, 
the cumulative distribution of received fluxes N($>$F). (We thank a
referee for pointing this out to us.)  When applied to a sample with known
fluxes, $V_{max}$ is interpreted as the maximum volume to which a specific
source could be seen given the flux limit of the detector. The 
probability distribution of $V_{max}$ contains information on the way 
in which the sources are distributed in space.  In particular, 
since $V/V_{max}$ is
proportional to $(F_{min}/F)^{3/2}$, the information in the distribution
of $V/V_{max}$ is identical to that contained in the N($>$F) distribution.
This was pointed out by Mao \& Paczy\'nski (1992) who give a simple, but
illuminating, analytic example of how the rate and peak luminosity can be
derived separately using the $V/V_{max}$ distribution for a given assumed
redshift distribution.  Thus derivations of GRB properties from
deviations of N($>$F) from an $F^{-3/2}$ form (e.g. Wijers et al. 1998;
Sethi \& Bargavi 2001; Stern, Tikhomirova \& Svennsson 2001) or
deviations in the distribution of $V/V_{max}$ from a uniform distribution
(e.g. Schmidt 1999, 2001) should be equivalent.  In practice, there
are differences.  Schmidt, Higdon, \& Hueter (1988) argue that the
$V/V_{max}$ statistic has advantages  because it is independent of
variations or fluctuations in background or instrumental sensitivity,
since the minimum flux and hence $V/V_{max}$ is recorded individually for
each object.  We have elected to adopt the results of Schmidt (1999)
as the basis of our work, in which, for a prescribed functional form
of the redshift distribution and luminosity function, the mean peak
luminosity and total GRB rate can be estimated.  Schmidt (2001) has
used a similar approach to actually derive the form of the luminosity
function as a superposition, which comes out to resemble the 
broken power law form
adopted in the earlier paper. (See, however, Sethi \& Bhargavi 2001,
who argue for a lognormal luminosity function).  

It should be emphasized that results such as we derive below, 
based on either method, must
assume the redshift dependence of the source density.  Yet another
method, suggested to us by a referee, would use only integrated surface
brightnesses of GRBs and blue stars. This method would be independent of 
the assumed redshift distribution if GRBs and high-mass stars are distributed 
identically in space, as we assume anyway, and if the redshift distribution 
of the two samples did not differ due to selection effects. This
approach can only give a product of the rate and mean peak
luminosity, which is insufficient for our applications, and is
subject to selection effects, as discussed in detail in Appendix A.

We recognize that these results will eventually be superceded
by better data and other methods, but it will be clear in what
follows how to modify our numerical results in such cases.  For
example, for an assumed form of the GRB peak luminosity function, its
parameters could be estimated from the small number of available
afterglow redshifts as a constraint added to the N($>$F) distribution
(Sethi \& Bhargavi 2001), or the luminosity function could be derived
by some proxy redshift indicator, such as the spectral lag-luminosity
relation (Norris, Marani, \& Bonnell 2000) or the variability-luminosity 
correlation (Fenimore \& Ramiriz-Ruiz 2001;
Reichart et al. 2001; Schaefer, Ding \& Band 2001).  Lloyd-Ronning,
Fryer, \& Ramirez-Ruiz (2001) have used the latter approach to
conclude that the luminosity function, whatever its form, is a
function of redshift, in contrast to what was assumed in previous
studies, in particular that of Schmidt (1999) on which our results are 
based.  Sethi \& Bhargavi (2001) have even argued that the redshift
distributions of the GRB and star formation rate differ.  
Clearly, there is more
to be learned about GRB properties, and the results we present here
are intended to be a reasonable representation of our current understanding.

In order to estimate the GRB rate in a given galaxy, say the
Milky Way, we avoid the practice of dividing the derived cosmic
rate by the total number of galaxies per unit volume derived
from the luminosity function, as in Wijers et al. (1998) and
elsewhere.  This division does give the rate per galaxy of mean
luminosity.  It does {\it not}  give the rate per
L*$_{\rm{gal}}$ galaxy as usually quoted; the mean
galaxy luminosity can be much
smaller than L*$_{\rm{gal}}$, depending on the slope and cutoff
of the low-luminosity portion of the adopted luminosity
function.  The mean number density diverges if
this slope is -1 or smaller.  Instead, we convert the GRB energy
production rate into a rate per unit stellar B-band luminosity.

Early estimates of GRB rates and mean peak luminosities
or energies assumed that all GRBs had the same luminosity,
the ``standard candle'' assumption (Wijers et al.~1998; Totani
1999; and earlier references given there).  The growing catalog of
GRBs with known redshift shows that L$_{\rm{peak}}$ spans three orders of
magnitude, even omitting SN~1999bw/GRB~990425, and the 
standard candle model has been abondoned (Krumholz, Thorsett,
\& Harrison 1998). 
To estimate the rate of production of GRBs, we draw on the analysis of
Schmidt (1999), who derived the local volume rate of GRBs and
associated properties.
Schmidt (1999) estimated the best fit parameters, including the local
GRB rate per unit volume and the characteristic peak luminosity,
for various assumptions about the redshift evolution of the bursts, 
the GRB peak luminosity function
(LF), taken to be a double power law with transition luminosity
L$_{\rm{br}}$ (denoted L* in Schmidt; we reserve L*$_{gal}$
for the luminosity of the break in the Schecter
galaxy luminosity function used below), and for two
choices of cosmological parameter q$_0$ (0.1 and 0.5).  Schmidt
found that a very narrow GRB LF, approximating the
standard candle model, cannot easily account for the large
redshift (3.4) of one of the observed GRBs with known
redshift (although Wijers et al. 1998 find a median redshift of
3.8 in their standard candle model).  Schmidt (2001)
finds direct evidence for a broken power law luminosity
function.  Schaefer, Deng \& Band (2001) also found a broken
power law from an
analysis of the proposed luminosity/variability relation
(Fenimore \& Ramirez 2001; Reichart et al. 2001)
and the lag-luminosity relation (Norris, Marani \& Bonnell 2000),
but, as noted above, Sethi \& Bhargavi (2001) argued for
a lognormal distribution and Lloyd-Ronning, Fryer, \& Ramirez-Ruiz (2001) 
argue that the distribution changes with redshift.  Krumholtz et al. 
(1988) showed that for models using GRB peak luminosity functions, 
the GRB data could not be
used to constrain the cosmic SFR history, since a broad range
of models were consistent with available constraints.  We adopt
this view here, that the SFR history must come from cosmological
observations of galaxies, but note that if the GRB luminosity-variability
and luminosity-lag relations are verified they give the
means to determine the history of the GRB formation rate directly
and hence, presumably, an independent means to determine
the SFR (Schaefer, Ding \& Band 2001).

Schmidt gives derived results for the case with no density
evolution (i.e.~constant GRB formation rate per comoving volume)
and for a strongly increasing
density $\propto$ (1+z)$^{3.3}$.  We refer to the latter case as
the ``strong evolution case;'' it is similar to the cosmic SFR
history to z=1 advocated by Madau et al. (1996) and others.
The results of Cowie, Songaila \& Barger (1999) for the
redshift dependence of the SFR from UV luminosity densities,
based on a very deep spectroscopic survey that overcomes
several problems with earlier UV work, give a much shallower
dependence, SFR $\propto$ (1+z)$^{1.5}$, so we have interpolated
between the various models examined by Schmidt in order to
allow for this weaker evolution of the SFR, which we refer to
as the ``intermediate evolution'' case.  As mentioned above,
the Guzman et al. (1997) [OII] study supports this case.  The
situation remains uncertain, however.  The H$\alpha$ luminosity
densities discussed by Yan et al. (1999) suggest that the
H$\alpha$ results might be consistent with the strong evolution
case, but this depends on the validity of the local H$\alpha$
luminosity density derived by Gallego et al. (1995).  The
Tresse \& Maddox (1998) H$\alpha$ result at z $\sim$ 0.3 is so
much larger than the local Gallego et al. value that it seems
possible that the local result is an underestimate.  The 
meta-analysis of the UV and H$\alpha$ results out to z = 1 by
Hogg (2001) favors the strong evolution case.  We will
quote results for the ``intermediate evolution'' case, but also
give the ``no evolution'' and ``strong evolution'' results for
comparison, and to show which results and conclusions are
sensitive to this uncertainty.  We assume that the derived
parameters depend only weakly on the assumed SFR history beyond
z=1, since the SFR is essentially unknown, but may be approximately
constant (see Pascarelle et al.~1998; Glazebrook et al.~1998;
Tresse \& Maddox 1998; Hughes et al.~1998), as assumed by
Schmidt (1999).  Note, however, that the SFR at high redshift
may still be plagued with problems of extinction.  Schaefer,
Ding \& Band argue that the GRB formation rate scales as
$(1+z)^{2.5\pm0.3}$, i.e. between our ``intermediate"
and ``strong" cases, for 0.2 $<$ z $<$ 5.

Schmidt tabulates a quantity
E$_{\rm{out}}$, which we will refer to as Q$_{\gamma,V}$
which refers to the GRB production per unit volume.  This quantity
is equal to the product of the local cosmic GRB rate r$_0$ (i.e., at z=0) in
units of Gpc$^{-3}$ yr$^{-1}$ (denoted $\rho$ by Schmidt) and
the {\it mean}  peak luminosity $<\rm{L_{peak}}>$
(in the 50--300 keV range), for each GRB LF
model, i.e. ${\rm Q_{\gamma,V} = r_0<L_{peak}>}$.
Note that the quantity called L* by Schmidt and
tabulated in his Table 1 is {\it not} the mean peak luminosity, but
the luminosity, L$_{\rm{br}}$,
at the break of the assumed double power-law
distribution function.  Schmidt's analysis thus allows
us to calculate $<\rm{L_{peak}}>$ as Q$_{\gamma,V}$/r$_0$.  We
note the dependence on the Hubble constant, H$_0$, of r$_0$,
$<\rm{L_{peak}}>$, and Q$_{\gamma,V}$ are H$_0^{~3}$, H$_0^{-2}$, and H$_0$.
In what follows, we adopt H$_0$ = 70~km s$^{-1}$ Mpc$^{-1}$, as in Schmidt.

Inspection of Schmidt's Table 1 shows that Q$_{\gamma,V}$ is rather
insensitive to the choice of cosmological parameter q$_0$ or
even the luminosity function model, but depends strongly on the
assumed redshift dependence of the GRB rate; for a GRB rate that
increases more rapidly with increasing redshift, the GRBs are
at larger average distance and must be brighter, but this is
outweighed by the much larger volume, reducing the derived
local rate.  For no evolution, Q$_{\gamma,V}$ $\approx$ 6 $\times$ 10$^{51}$
erg s$^{-1}$ Gpc$^{-3}$ yr$^{-1}$, while for the (1+z)$^{3.3}$
evolution, Q$_{\gamma,V}$ is between 9.0 $\times$ 10$^{50}$ and 1.5 $\times$
10$^{51}$ erg s$^{-1}$ Gpc$^{-3}$ yr$^{-1}$ for the various LF
models tabulated by Schmidt, and we adopt 1.3 x 10$^{51}$ erg
s$^{-1}$ Gpc$^{-3}$ yr$^{-1}$ in that case. Following Schmidt
(1999), we have multiplied the tabulated values of Q$_{\gamma,V}$ by a
factor of 2.1 to convert from the energy radiated in the 50--300
keV range to that in the 10--1000 keV range.  Bearing in mind
the Cowie et al. (1999) result, which gives an increase in SFR
of 2.8 out to z=1 instead of 9.8 for the strong evolution case,
and assuming that the GRBs follow the SFR, we adopt a value of
Q$_{\gamma,V}$ for the intermediate evolution case as the average of the
no-evolution and strong evolution cases, giving Q$_{\gamma,V}$ = 3.8
$\times$ 10$^{51}$ erg s$^{-1}$ Gpc$^{-3}$ yr$^{-1}$ (the odd time units
will be rationalized below when we multiply by the duration of
a GRB so that the erg s$^{-1}$ becomes a total energy emitted in
a mean burst), with an
uncertainty of a factor of two.  We realize that there is no
best way to interpolate the Cowie et al. evolution between the
other two cases, and have chosen to simply adopt the average.
Note that the units of erg s$^{-1}$ that are
associated with Q$_{\gamma,V}$ represent a ``marker" for the
place of the GRB in the peak luminosity distribution function.

We take a blue luminosity density of
\begin{equation}
\rm{J_{gal,B,V} \approx 1.5 \times 10^8~h_{70}~L_{\odot}~Mpc^{-3}}
= 6.0\times10^{41}~erg~s^{-1}~h_{70}~Mpc^{-3}
= 6.0\times10^{50}~erg~s^{-1}~h_{70}~Gpc^{-3} ,
\end{equation}
from the ESO Slice Project result of Zucca et al. (1997).
A revised estimate reaching lower surface brightness
from Cross et al. (2001) is
$\rm{J_{gal,B,V} = 1.75 \times 10^8~h_{70}~L_{\odot}~Mpc^{-3}}$, so
our adopted number is probably accurate to within 20 percent.
It is important to use a deep survey because we are comparing with
GRBs that are seen out to very large redshifts.
The uncertainties even in the deep samples include differences in
methodologies, systematic photometric errors, partial exclusion of
low surface brightness galaxies, and other effects, as reviewed by
Loveday (1999). 
  A thorough discussion of the galaxy LF is given by
Marinoni et al. (1999).
We can combine the rate per unit volume of GRBs with this value of
the blue luminosity density to estimate the rate of production
of GRBs per unit blue luminosity, ${\rm r_B = r_0/J_{gal,B,V}
= 4.4, 2.4, and~0.3\times10^{17}~h_{70}^2~\l^{-1}~yr^{-1}}$, for no,
intermediate, and strong SFR evolution, respectively.

>From Binney \& Merrifield (1998), we take a mass
surface density in the solar neighborhood $\Sigma_M$ = 45
M$_{\odot}$ pc$^{-2}$, and a local value of M/L$_{\rm{B}}$ =
2.3 \m/\l, giving the blue luminosity surface density
$\Sigma_{\rm{L,B}}$ = 20 L$_{\odot}$ pc$^{-2}$ = $8\times10^{34}$
erg s$^{-1}$ pc$^{-2}$.
We can then form the product (which is independent of H$_0$)
\begin{equation}
S_{\rm{MW}} <L_{\rm peak }> = \frac{Q_{\gamma,V}}{J_{gal,B, V}}
 \Sigma_{\rm{L,B}} =
5.1\times10^{35}~{\rm ~erg~s^{-1}~pc^{-2}~yr^{-1}},
\end{equation}
for the case of intermediate evolution, where $S_{MW}$ is the rate of
occurrence of
GRBs per unit area, the subscript MW refers to the Milky Way, and we remind
the
reader that the  units of erg s$^{-1}$ here refer to the peak power in an
individual
GRB and not, for instance, some time-averaged power integrated over the
Galaxy
(see Appendix A).  The utility of the product $\rm S_{MW}<L_{peak}>$ is
described below.

We tabulate various useful quantities in Table 1 for our
three choices for the evolution of GRBs,
noting that we take the intermediate
evolution result as the most realistic case.  The adopted
values are for q$_0$ = 0.5.  For q$_0$ = 0.1, Q$_{\gamma,V}$ is relatively
unaffected, while r$_0$ is smaller by a factor of about 0.4,
$\rm{<L_{peak}}>$ will be larger by a factor of 2.4, and
S$_{\rm{MW}}$ is smaller by a factor of 0.4.  It can be
seen from Eqn. B8 in Appendix B that the
probabilities of occurence and average times between events for a
given received fluence depend only on the quantity
\begin{equation}
\label{FV}
F_V = S_{\rm{MW}} <L_{\rm peak}>\Delta t_{\gamma} =
5.1\times10^{36}\left(\frac{\Delta t_{\gamma}}{10~s}\right) ~{\rm
erg~pc^{-2}~yr^{-1}}
=  1.6\times10^{29}\left(\frac{\Delta t_{\gamma}}{10~s}\right) ~{\rm
erg~pc^{-2}~s^{-1}},
\end{equation}
where $\rm{\Delta t_\gamma}$ is the average GRB peak duration,
taken to be 10 sec which is the average fluence/peak flux ratio found by
Schmidt
(1999).  The quantity $F_V$ is insensitive to q$_0$.
This energy production rate per
unit area can be scaled to other galaxies by multiplying by
$\rm{L_B/L_{B,MW}}$, assuming that $L_{\rm{B}}$ measures the
recent SFR and the GRB rate.

For their assumption of strong stellar evolution with redshift 
(roughly comparable to ours), the standard candle peak luminosities 
derived by Wijers et al. (1998) are about a factor of three larger 
than the mean  ({\it not} L$_{\rm{br}}$) peak luminosities we obtain 
from Schmidt for the strong evolution case.  For no evolution, their
peak luminosity is a factor of two smaller than ours.  The use of a 
LF has reduced the dependence of mean peak luminosity on evolution 
model from a factor of nearly 20 to a factor of only three, as can
also be seen from Schmidt's examples.  On the other hand, the
standard candle cosmic rates (Gpc$^{-3}$ yr$^{-1}$) are a factor
of three smaller in both cases, an effect which is not seen in
Schmidt's standard candle cases.  The Schmidt results refer to
the 10--1000 keV range, while Wijers et al. use the 30--2000 keV
range, but this should not contribute much to the difference in
results.  

Porciani \& Madau (2001) have also estimated GRB rates and associated
properties for three adopted star formation rate prescriptions.
They find larger GRB peak luminosities and smaller local
production rates by factors of several compared to Schmidt (2001) who
adopted the same three star formation rate prescriptions.  Schmidt also 
finds the local GRB rates to vary with respect 
to the star formation prescriptions 
in the opposite sense than do Porciani \& Madau.  Up to a redshift
of 1 these three prescriptions all correspond roughly to our strong evolution
case and the results derived by Schmidt correspond closely with those
given here for that case.  The differences between Porciani \& Madau and
Schmidt may be due to the use by Porciani \& Madau of a 
single power law rather than
a broken power law luminosity distribution as derived by Schmidt (2001)
and used by Schmidt (1999), the basis of the current analysis.
Our rates are larger by about a factor of 10 than estimated for the
strong evolution case by Dar \& De R\'ujula (2001).

The GRB rates and related quantities presented in this section are 
``isotropic equivalent" rates and have not been corrected for collimation.
Some applications require this correction, but others do not.
We next use these derived quantities to examine some
astrophysical and astrobiological implications.
We describe the effects of collimation where relevant.

\section{GRBs AND SUPERNOVAE}

In the context of models in which GRBs arise in
massive stars, it is interesting to consider the rate of
occurence of GRBs in comparison to SN.
Supernova rates are typically given in units of number per
10$^{10}$ L$_{\odot}$ of luminosity in the blue band per century,
known as a ``supernova unit'' or SNu.  The rates per unit
blue luminosity for GRBs, ${\rm r_B}$, can be cast in these units to give
4.4, 2.4 and 0.3$\times$ 10$^{-5}$ h$_{70}^{~~2}$ SNu for 
no, intermediate, and strong evolution respectively.
For a blue luminosity of the Milky Way of $2.3\times10^{10}{\rm \l}$
(Trimble 2000),
the total rate of GRBs in the Galaxy would be
10, 5.5, and 0.7 $\times10^{-7}$ yr$^{-1}$, respectively.

Cappellaro et al. (1997) give rates for Type Ib/c and Type II
supernovae (SN Ib/c; SN II)
as a function of galaxy type.  The rate of SN Ib/c
averaged over galaxy types Sbc-Sd is 0.14 $\pm$ 0.07
h$_{70}^{~~2}$ SNu.  The rate for SN II is about a factor of
5 higher with similar uncertainty.  The ratios of the rate of
SN Ib/c supernovae in Sbc-Sd galaxies compared to the rate of
GRBs are 3,200, 5,800, and 46,000 for no,
intermediate, and strong star formation evolution,
respectively.  These estimates are smaller by a
factor of 10 - 100 than those given by Lamb (1999).
Our ratios would be higher by about a factor of
3 if we considered SN rates in Sc galaxies alone.
There have been suggestions of correlations of GRBs
with some SN II (Germany et al.~1999), although
physical constraints associated with the thick hydrogen
envelope suggest that it would be difficult to generate the
requisite relativistic flows in that environment (MacFadyen \&
Woosley 1999).  If GRBs were to be associated with
SN II, the ratios of rates would be higher by a factor of about 5. 
Porciani \& Madau (2001) assume SN~II come from all stars above 
8 \m\ and derive a rate of about $5\times10^5 - 10^6$ SN II per
GRB, a factor of 2 - 4 larger than our rate for the strong evolution case
and about 20 larger than our intermediate evolution case.
As noted above (\S 2), Porciani \& Madau give a GRB rate that is 
smaller than that of Schmidt (1999, 2001), so this could account 
for the difference in the strong evolution case.  
These ratios of supernovae to GRBs could all be reduced by a
factor of $\Delta \Omega /4 \pi$ if all GRBs are
collimated by the same universal amount.  Although
their analyses differ in substantial ways, both Panaitescu \& Kumar (2001)
and Frail et al. (2001) suggest that rather tight collimation may
be the rule rather than the exception with $\Delta \Omega / 4 \pi$
ranging from 0.001 to $\gta$ 0.01.

The ratio of SN to GRB rates is relevant
to a variety of astrophysical issues and to the nature of the
GRBs themselves.  In particular, this ratio represents
constraints on the currently popular picture that
GRBs are associated with star formation (the ansatz behind
Schmidt's calculation) and hence that GRBs arise in the
collapse of massive stars to produce black holes (MacFadyen \&
Woosley 1999) or rapidly spinning neutron stars (Wheeler et al.~2000).  If
collimation is not a significant factor in the mean
rates derived by Schmidt, but only, for instance, in the rare
bursts with exceptionally high isotropic equivalent energy,
then GRBs must be extremely  rare.  If, for
instance, they only come from stars more massive than a given
threshold value, then that threshold must be exceedingly high.

As an illustration, if the integrated number of stars with mass above
some value, M, scales as M$^{-n}$, GRBs occur for
all stars above a threshold, M$_{\rm{GRB>}}$, and SN
occur in stars with mass in excess of M$_{\rm{SN>}}$, then
M$_{\rm{GRB>}}$ = M$_{\rm{SN>}}$
(N$_{\rm{SN}}$/N$_{\rm{GRB}}$)$^{\rm{1/n}}$.   Note that the
actual value of the supernova threshold is not too important
since it just enters linearly; we take 10 M$_{\odot}$ as a
representative value.  The value of  n is 1.3 for a Salpeter
mass function and n = 1.8 represents a steeper,
but still reasonable, mass function for massive stars (see Scalo 1998 for a
critical review of cluster IMFs).  For the ratios given above
and a Salpeter slope, GRBs could only occur for stars with a mass in
excess of M$_{\rm{GRB>}}$ = 5000, 8000, and 40,000 M$_{\odot}$,
for no, intermediate, and strong SFR evolution, respectively.
For a  steep slope, n = 1.8, the numbers would be
M$_{\rm{GRB>}}$ = 900, 1200, and 4000 M$_{\odot}$, respectively.
If SN II rates were adopted, these mass limits would all be
increased by a factor of 2--4, depending on the slope of the
mass function.  Clearly, if collimation or some other effect
does not significantly alter the rates given by Schmidt,
GRBs cannot arise by ``normal'' black hole
formation, nor can they be driven by the birth of every
magnetar, which might represent a fraction of order 10 percent of
``normal'' pulsars.

If the majority of \grbs\ are significantly collimated, then
these estimates will change.  For the
assumption that all GRBs are collimated by a factor
of $\Delta \Omega / 4 \pi$ =  0.01, the ratios of SN Ib/c to
GRBs would give threshold values of M$_{\rm{GRB>}}$
= 140, 230, and 1100 \m\ for n = 1.3 and 69, 96, and 300 \m\ for n =
1.8, respectively.  If values of $\Delta \Omega / 4 \pi =  0.001$
were to prove typical, then the corresponding mass limits would
be 25, 39, and 190 \m\ for n = 1.3 and 19, 27, and 84 \m\ for
n = 1.8.  These limits are in the range of ``normal" supernovae,
suggesting that a substantial fraction of all SN Ib/c could
produce \grbs.  The scanty statistics imply that this
issue is still open, but that for collimation factors in the
range 0.001 to 0.01, GRBs could arise from ``normal" massive stars,
depending on the slope of the IMF.

We note the caveat
that, while SN Ib/c are thought to be associated with massive
stars, their progenitor evolution is unknown, so assigning a
minimum mass to GRBs on the basis of rates may not
be entirely appropriate.  In particular, if SN Ib/c require
binary evolution and mass transfer then they do not sample the
initial mass function in any straightforward way.  Given the
empirical rates, however, it is clear that even with
substantial collimation of GRBs, there are large
ranges of reasonable parameter space where GRBs
must represent extremely massive progenitors or otherwise
select a small portion of the total mass range available to SN
Ib/c progenitors.  

The results for the combination of parameters that seem most
reasonable to us, namely intermediate redshift evolution,
an average SN rate for Sbc-Sd galaxies, only SN Ib/c (not SN II) being
associated with GRBs, and in addition using a relatively steep
IMF index for massive stars of n = 1.8 (steeper than most claims in the
literature) gives M$_{\rm{GRB>}} \sim 30$ M$_{\odot}$
only if the average collimation is $\Delta \Omega / 4 \pi \sim 0.001$.
Thirty solar masses is a plausible lower mass
limit for black hole formation (Twarog \& Wheeler
1982; Fryer 1999).  This means that GRBs might be associated
with routine black hole formation.  The situation becomes more extreme
if one or more of the following obtain:  less collimation,
strong cosmic evolution, comparison is made to SN II supernova rates,
or an IMF as flat as n = 1.3.  In such cases, routine black hole
formation may not be able to play a role in producing GRBs.
It may be that only
special cases with exceptionally high initial stellar rotation
or magnetic field can generate a GRB.  The same
statement applies to models based on rapidly rotating, highly
magnetized neutron stars.  Only a tiny fraction of such events,
perhaps again those with exceptionally strong rotation and
magnetic field, could contribute GRBs.  If magnetars represent
about 10 percent of all core collapse supernovae, then every magnetar
birth could, statistically, generate a GRB if again the
collimation falls in
the range  $\Delta \Omega / 4 \pi \sim  0.001$ for the
``reasonable" parameters defined above.

Finally, we note that, despite their large energy, the low
rate of GRBs compared to supernovae means that GRBs are
unlikely to have a significant impact on induced star formation
or on nucleosynthesis except, perhaps, for some rare species that
might be specifically produced in GRBs.

\section{GRBs AND HI HOLES IN GALAXIES}

It has been long known that very large, shell-like structures
exist in the HI distribution of the Milky Way (Heiles 1979) and
other galaxies (Brinks 1981), from large disk galaxies like
M101 to dwarf galaxies like the SMC and IC 2574 (see Wilcots \&
Miller 1998; Staveley-Smith et al.~1997; Kim et al.~1998;
Walter \& Brinks 1999, and references therein;  see Walter
1999 for a review).  A typical galaxy has 50--500 shells/holes
with sizes in the range 0.1--1 kpc and typical ages of 10$^7$
yr.  The possible processes for producing such structures have
been extensively discussed (e.g.~Tenorio-Tagle \& Bodenheimer
1988; Walter 1999), with a leading candidate being winds driven
from young clusters by OB star winds and multiple supernovae.
Such a model successfully accounts for the size distribution of
hole sizes in the SMC (Oey \& Clarke 1997), although there are
problems for other galaxies, and possibly for the assumed
importance of stalled shells in that work (Walter \& Brinks
1999).  There may also be a problem with the energy required to
account for the largest holes, although this might require a
different process for only a small percentage of the holes.
Rhode, Salzer, \& Westpfhal (1999) failed to detect
the expected residual populations of the putative OB
associations responsible for the holes in the dwarf galaxy Ho
II, leading them to suggest some other mechanism is required
(see however the cautionary remarks in Efremov, Elmegreen, \&
Hodge 1998 and Walter \& Brinks 1999).

Efremov et al. (1998)
and Loeb \& Perna (1998) have independently suggested that GRBs
could be the primary process responsible for the HI holes (see
also Efremov 1999a,b, 2000 for arguments specifically aimed at
stellar arcs and a supershell in the LMC).  It is therefore of
interest to examine the viability of GRBs as an explanation for
most of the HI holes in light of the values for GRB peak
luminosities and galactic rates we have inferred from Schmidt
(1999) and the reduction in SFR redshift evolution based on
Cowie et al. (1999).  We perceive two problems with associating GRBs with
the large
HI holes: the energy required to drive them and the number of such holes
in galaxies of small luminosity and hence small expected GRB rates.
One can have one or the other, but not both.

The large HI holes require a large energy input which is why
they are traditionally associated with the effects of 10 - 100
supernovae and their progenitor stars.  For GRBs, the required
energies are produced only if the conversion of kinetic energy
to \gr\ energy is rather inefficient and if collimation is
neglegible for most \grbs.  As an illustration, our best choice value of
$<\rm{L}_{\rm{peak}}>$ is 1.1 $\times$ 10$^{51}$ erg s$^{-1}$.
Using the fluence/flux ratio of 10 sec estimated by Schmidt, and
assuming an efficiency for conversion of kinetic energy to
radiation of $\epsilon = 0.01$, as assumed by Efremov et al. (1998) and
Loeb \& Perna (1998) and supported by arguments given by Kumar
(1999), we obtain an average isotropic equivalent kinetic
energy of E = 1.1 $\times$ 10$^{54}(\epsilon/0.01)^{-1}$ erg.
Use of  the same late phase
blast wave scaling relation as employed by Efremov et al. and Loeb
\& Perna (Chevalier 1974) shows that a typical shell should
slow down to 10 km~s$^{-1}$ at a radius R$_{\rm{kpc}}$ = 0.7
E$_{54}^{~~0.32}$n$^{-0.36}$, where n is the ambient gas number
density, assumed uniform.  We take n = 1 cm$^{-3}$, although there
is some evidence that the average particle density may be
somewhat smaller in the ``puffed up'' dwarf IC 2574 and other
dwarf galaxies (see Walter \& Brinks 1999).  The time at which
this radius is reached is t = 20 E$_{54}^{~~0.32}$n$^{-0.36}$
Myr.  Since the isotropic equivalent energy found here is very
similar to the total actual energy
assumed by Efremov et al. (1998) and Loeb \& Perna (1998), we
agree with their conclusion that GRBs could account for the sizes
of shells and their estimated ages using the adopted
parameters. We note that the value of $\epsilon$ is
controversial; if $\epsilon$ is as large as 0.85, as claimed by
Fenimore \& Ramirez-Ruiz (2000), the effectiveness of GRBs for
explaining galactic HI holes will be compromised with respect
to accounting for the observed sizes and ages, even before
accounting for collimation effects.
Collimation effects are, however, critical for this argument.
If collimation with $\Delta \Omega / 4 \pi \sim$ 0.001 - 0.01 is
the rule, as discussed in \S 3, and typical explosion energies
are $\sim 5\times10^{50}$ ergs (Panaitescu \& Kumar 2001; Frail et al.
2001),
then it is difficult to see how GRB could contribute to any of
the large HI holes.  With the scaling for size given above, a reduction
of the energy by a factor of $\sim 1000$ would yield holes only
100 pc, not 1 kpc, in size.

Interestingly, substantial collimation could
ameliorate a problem with the number of HI holes in small galaxies.
The GRB hypothesis as proposed by Efremov (1998) and by Loeb \& Perna (1998)
fails to account for the
large number of holes observed in many dwarf galaxies, even if
large energies are assumed.  Following Loeb \& Perna, the
average number of holes observed at any time should be
approximately equal to the ratio of the mean shell age (as
given above) to the mean time between GRB, which is the inverse
of the GRB rate.  The masses of many of the galaxies in which
numerous holes are found are much smaller than the Milky Way,
so the GRB rates uncorrected for collimation are much smaller
than estimated for the Milky Way,
leading to large times between events and therefore an
unacceptably small prediction for the number of shells.  For
example, consider the M81 Group dwarf IC 2574 studied by Walter
\& Brinks (1999), the luminosity of which
is L$_{\rm{B}}$ = 8 $\times$
10$^{8}$ L$_{\odot}$.  Using our best estimate for the GRB rate
per unit blue luminosity, 
${\rm r_B = 2.4\times10^{-17}~\l~yr^{-1}}$, the mean
GRB rate should be 1.9 $\times$ 10$^{-8}$ yr$^{-1}$,
corresponding to a mean time between events of 50 Myr.  Thus
the probability of observing even one shell of size and age
given above is less than unity for this galaxy, while at least
50 holes are observed by Walter \& Brinks.  No reasonable
decrease in the assumed density could yield agreement.
Collimation with $\Delta \Omega / 4 \pi \sim$ 0.001 - 0.01 would
bring agreement with the rates, but with correspondingly small energy.
A similar disparity occurs for other dwarf galaxies, including Ho
II (Puche et al.~1992; Rhode et al.~1999) and the SMC
(Staveley-Smith et al.~1997), where large numbers of large holes
are observed even though these galaxies are somewhat fainter
than IC 2574.  Even in the Local Group dIrr galaxy IC 10, where only
7--8 HI shells were found by Wilcots \& Miller (1998), the
discrepancy remains large, since the luminosity of IC 10 is
only L$_{\rm{B}} \approx 2.4 \times 10^{8}$ L$_{\odot}$.  The
expected GRB rate is more than three times smaller than given
above for IC 2574, and the probability of even a single hole is
smaller by the same factor.  The radii of all the shells in
IC 10 are only around 50 pc, requiring small explosion energies
and the discrepancy is somewhat less in this case. This galaxy might
thus be consistent with the frequent, smallish holes that could
be produced by collimated GRBs, but even here the need for GRBs is
neither compelling nor unique.
Note that if the case of maximum evolution (1 + z)$^{3.3}$ were
adopted, the situation becomes much worse, because the average
cosmic rate, and hence specific rate per unit mass or
luminosity, decreases by a factor of seven.

We thus find that GRBs cannot simultaneously solve the constraints of
the large apparent energy required to generate typical large HI holes
and their large numbers in small galaxies.  If the GRB energies are as
large as adopted by Loeb \& Perna (1998), collimation must be
negligible and the rate of GRBs must be correspondingly low.  Our
difference with Loeb \& Perna in this regard is
therefore primarily due to these authors neglecting to notice
the small masses of many of the galaxies with numerous holes.
Models based on winds driven by multiple SN do not
suffer from this disparity because, even though 100--1000 SNe
may be required to explain the large holes, the rate of SNe is
many orders of magnitude larger than the rate for
GRBs if the latter emit isotropically.
Similarly, we find that the statement by Efremov et al. (1998)
and Efremov (1999a,b)
that 4--5 very large HI shells or arcs in
the LMC could have been produced by GRBs over the past 10$^{7}$
yr is untenable, especially considering the relatively small
mass of the LMC and its present SFR and the large energy requirements.
Our analysis is based on long
GRBs and Efremov invokes binary neutron star mergers that might not
produce long bursts, so our constraints might not apply directly.

It is still possible that a small number (of order unity) of
small ($\lta$ 100 pc) holes per galaxy could be due to GRBs, if most
of the explosions
are initially highly collimated.  At the large sizes and
ages at which the GRB explosions would be observed as large HI
holes, the collimation could have decreased considerably.  It is also
possible that
GRBs could account for rare, high energy remnants.  The largest HI holes
$\rm{(\sim~1.4~kpc}$ in radius),  require an  energy of about 8 $\times
10^{54}$ erg.
The distribution of GRB energies is likely a decreasing
function of energy with few, if any, reaching such energies.
For example, using a peak luminosity
function N(L) $\sim$ L$^{-1.5}$, a slope between the two
high-luminosity power law slopes adopted by Schmidt (1999),
energies E = L$\Delta$t$_\gamma$ this large are expected in
only a fraction $\sim~5\times10^{-5}$ of events.
GRBs thus might still account for the most luminous remnants,
for example the hypernova candidates in M101, if the
energy is not severely collimated.  Contrasting views are given
by Wang (1999) and Lai et al. (2001).  Hydrodynamic
simulations can probably help resolve the question of the
association of GRBs with the M101 hypernova remnants (Kim, Mac
Low \& Chu 1999).  Whatever the resolution of this question, we
only claim to have shown that the GRB rate is far too small to
account for most of the large HI holes in galaxies.

\section{GRBs AND CHONDRULES}

Chondrules are submillimeter-sized meteoritic silicate
inclusions that appear to have solidified by cooling after
rapid heating during the first $10^7$ yr  of the history of the
solar system.  The nature of the heating process that melted
the chondrules in the early solar system has remained enigmatic
for many years.  A detailed review of the empirical constraints
and most of the proposed heating models has been given by Jones
et al. (2000; see also Cohen, Hewin \& Yu 2000).
McBreen \& Hanlon (1999) have made the
intriguing suggestion that GRBs could supply the fluence
required to melt the chondrules.  One piece of evidence in
favor of this idea is that the textures and other properties of
chondrules imply that the heating event was short lived,
probably less than a minute, in agreement with the ten second
average GRB duration.  Since McBreen \& Hanlon assumed a very
large GRB energy, and because we are unable to reproduce the
rates that they adopted, it is of interest to re-examine the question.

McBreen \& Hanlon (1999) estimate that $2 \times 10^{10}$ erg
g$^{-1}$ is needed for the melting of chondrule precursors and
calculate the minimum GRB fluence required to produce chondrule
layers of thickness 0.18, 0.8, and 2 g cm$^{-2}$ as $1.8 \times
10^{10}$, $7.0 \times 10^{10}$, and $1.5 \times 10^{11}$ erg
cm$^{-2}$.  We adopt these values in the present calculations.

We need to estimate the probability p(F$_{\rm{cr}}$) that such a
fluence  occurred during the first 10$^7$ yr of the life of the
solar nebula.  For a two-dimensional Poisson process the result
given in Appendix B (Eqn. B5) yields
\begin{equation}
\rm{p(F_{cr}) = 
1 - exp 
\left[-\frac{\pi}{4}\left(\frac{\ell_{cr}}{\bar{\ell}}\right)^2\right]},
\end{equation}
\noindent where $\ell_{\rm{cr}}$ is the critical distance at which
a GRB of fluence ${\rm <L_{peak}>\Delta t_\gamma}$
can produce a fluence F$_{\rm{cr}}$
at the Sun (Eqn. B7):
\begin{equation}
\rm{\ell_{cr} = 
\left(\frac{<L_{peak}> \Delta t_\gamma}{4 \pi F_{cr}}\right)^{1/2}},
\end{equation}
and $\bar{\ell}$ is the mean distance expected for GRB
markers within elapsed time t, in this case $10^7$ yr,
given the rate per unit area of the events, ${\rm S_{MW}}$ (Eqn. B4),
\begin{equation}
\rm{\bar{\ell} = \frac{1}{2}(S_{MW} t)^{-1/2}}.
\end{equation}
  Using our best estimates for the mean GRB
fluences, ${\rm <L_{peak}>\Delta t_\gamma}$ = $1.1\times 10^{52}$ ergs,
we find that the three values of F$_{\rm{cr}}$ given
above correspond to $\ell_{\rm{cr}}$ = 71, 36, and 25 pc.
Note that this result is independent of the collimation
and isotropic equivalent energy.  For the GRB rate per unit area in the
solar
vicinity at the time of solar system formation, we follow
McBreen \& Hanlon (1999) and allow for the possibility that
the Sun was formed in a spiral arm, where the SFR may be
larger.  Whether spiral arm passage actually enhances the SFR
per unit gas mass, or instead simply increases the gas density
and hence only the SFR per unit volume, is still a contentious
issue.  A detailed discussion of various considerations is
given in Elmegreen (1997).  Based on these considerations, we
increase the average GRB rate per unit area that we derived for
the Milky Way by a factor of three to account for the
spiral arm effect.  This gives $\bar{\ell} = 4200~\rm{pc~(t/10^7
~yr)^{-1/2}}$.  For the three critical fluences, the
probability p(F$_{\rm{cr}}$) is $2.2 \times 10^{-4}$, $5.8
\times 10^{-5}$, and $2.8 \times 10^{-5}$.
Rather than {\it assume} that GRBs formed the chondrules and
then conclude that chondrules are very improbable for {\it other}
planetary systems as suggested by McBreen \& Hanlon (1999), 
it seems more reasonable to us to conclude
that GRBs are an improbable source of heat for solar system chondrules.

The situation is actually more pessimistic than this.  McBreen
\& Hanlon (1999) fail to consider the strong evidence that a
significant fraction, if not most, chondrules, have experienced
more than one heating event (Rubin \& Krot 1996),  although
they point out the possibility without further discussion.
For example, many formations consist of chondrules within larger
chondrules.  We can estimate the probability that the primitive
solar nebula was subjected to two heating events during its
first 10$^7$ yr from the Poisson spatial model (Eqn. B1), for
which the probability of two events within area A during time t
is
\begin{equation}
\rm{p(2) = \frac{1}{2}(S_{MW} tA)^2~exp(-S_{MW} tA)}.
\end{equation}
This expression is
only valid as long as $\ell_{\rm{cr}}/\bar{\ell}$ is not so large
that three or more events have occurred, a condition easily
satisfied in the present case.  Using $\rm{A = \pi
\ell_{cr}^{~~2}}$ and $\rm{S_{MW} t = 1/4\bar{\ell}^2}$, we get
\begin{equation}
\rm{p(2) = \frac{1}{2} \left[\frac{\pi}{4}
\left(\frac{\ell_{cr}}{\bar{\ell}}\right)^2\right]^2~
exp\left[-\frac{\pi}{4}\left(\frac{\ell_{cr}}{\bar{\ell}}\right)^2\right]}.
\end{equation}
Using our best estimate for the rate and mean fluence and the
three sample critical fluences adopted by McBreen \& Hanlon, we
find p(2) = $2.5 \times 10^{-8}$, $1.7 \times 10^{-9}$, and
$3.9 \times 10^{-10}$, respectively.  Thus it is clearly
unlikely that GRBs contributed to the multiple heating of the
chrondrules.

\section{GRBs AND BIOLOGICAL EVOLUTION}

        In this section we are interested in the following question: 
What is the probability per unit time that the Earth or other 
habitable planets in our Galaxy have been irradiated by a GRB fluence 
capable of causing significant biological affect, either through 
direct DNA alterations of surface organisms or by longer-lived 
alterations in the chemical makeup of the atmosphere?   We are 
interested in stochastic effects that could affect biological 
evolution at any level and are not specifically concerned with 
catastrophes like mass extinctions induced either directly or through 
changes in atmospheric chemistry, although the latter may indeed be 
important; see Ruderman (1974), Crutzen \& Bruhl (1996), Collar 
(1996), Ellis, Fields \& Schramm (1996), Thorsett  (1995), Dar, Laor \& 
Shaviv (1998) and Dar \& DeR\'ujula (2001) for a variety of discussions 
focusing mainly on supernova explosions and for earlier references.  
Only the latter three papers discussed GRBs, but they were concerned 
with ozone layer destruction, production of local radioactive species, 
and showers of atmospheric muons that can penetrate underground and underwater.  
We concentrate here on direct biological effects due to DNA damage.
We do not consider the possible production of TeV \grs\ (Dar \& DeR\'ujula)
but if such radiation exists at large fluence levels, it is 
undoubtedly important.
    
In order to estimate the frequency of biologically 
significant events it is necessary to understand both the nature of 
\gr\ transport through a planetary atmosphere and the critical 
fluences of both ionizing radiation (for thin atmospheres) and UV 
radiation (for thick atmospheres, see below) necessary for direct 
biological affect.

\subsection{Radiation Transport} 

The energy at which $\nu~F_{\nu}$ peaks in most GRBs is about 200 keV with 
a tail in the distribution of peak energy extending to $\sim$ 1 MeV (Band 
et al. 1993; see Preece et al. 2000).  The passage of a beam of such 
photons through a planetary atmosphere will result in the following 
sequence of events.  The incident high-energy photons will be Compton 
scattered to lower energies until, after 2-5 scatterings (or escaping 
the atmosphere by backscattering) 
the energy reaches $\sim$ 50 to 100 keV.  Below this energy, 
X-ray photoabsorption dominates Compton scattering for any 
composition of interest for planetary atmospheres and the atmosphere 
is ``black" all the way to the far ultraviolet, where N$_2$ and CO$_2$ (if 
present in large abundance) maintain  a large opacity out to about 
100 nm or 200 nm respectively.  Monte-Carlo simulations of this 
process for a range of column density and other parameters (Smith, 
Scalo \& Wheeler 2001) show that the fraction of photons above this 
energy that reach the ground is exponential in the column density, 
with an e-folding  column density of about 35 g cm$^{-2}$, with most of 
the photoabsorption occurring at an altitude of 5-20 km, depending on 
the column density, for a terrestrial gravity.  

This result 
suggests that we consider two different classes of atmospheres. 
``Thin" atmospheres with column densities less than about 100 g 
cm$^{-2}$ will receive a significant fraction of the incident radiation 
in the form of ionizing radiation at the planetary surface (spectra 
are discussed in Smith et al. 2001).  In these cases we must estimate 
the threshold fluence required for biological damage due to ionizing 
radiation.
In the case of ``thick" atmospheres with column densities greater than 
about  300 g cm$^{-2}$, like the Earth (1050 g cm$^{-2}$), the fraction of 
radiation that reaches the ground at X-ray energies or higher is 
negligible; however, there can still be a large fraction of energy 
reaching the ground in biologically-interesting ultraviolet radiation 
because of the importance of secondary effects.  

Each Compton scattering (high in the atmosphere) or X-ray 
photoabsorption (lower in the atmosphere) 
results in a relatively high-energy primary 
electron that loses energy through collisional ionization.  The 
secondary electrons so released will have a range of energies, but a 
typical average energy per ionization event is 10 to 40 eV.  These 
secondary electrons must eventually lose their energy through 
excitation of electronic levels (possibly after further ionizations) 
in the major atmospheric constituents (generally C, N, O), and 
through elastic (Coulomb) interactions.   The excitation losses 
result in radiative deexcitation and the emission of photons from the 
far UV to the infrared, depending on the atoms and levels involved. 
For example, a 1 MeV incident photon suffering a Compton scattering 
will produce a recoil electron with an average energy of about 
500 keV, and nearly all of this energy will be lost along its path by 
collisional ionizations, producing more than $10^{4}$ secondary 
electrons that will, through excitation, redistribute the original 
500 keV in the form of spectral lines.  To the extent that these 
spectral lines occur in the UV, specifically in the regions that can 
cause significant biological affect ($<$ 300 nm) but still reach the 
ground because the atmospheric opacity is small enough ($>$ 200 nm), a 
significant fraction of the original GRB energy fluence can thereby 
reach the ground as UV radiation capable of DNA alterations (the DNA 
photoabsorption cross section peaks strongly at 260 nm).  Preliminary 
radiative transfer calculations including secondary effects and using 
simplified model carbon and oxygen atoms as the dominant atmospheric 
constituent for the level population calculations (H\"oflich 2001, 
private communication) indicate that about 1 percent of the incident energy 
can reach the ground in the UV for an atmosphere with the Earth's 
column density.  We adopt this value in what follows although we 
realize that more detailed calculations are required to further 
quantify this number and its dependence on parameters. 

As far as atmospheric chemistry is concerned, the distinction between 
``thin" and ``thick" atmospheres is less distinct, because we find 
that, even in the case of ``thick" atmospheres like the Earth's, the 
bulk of X-ray photoabsorption occurs at relatively low altitudes (for 
a terrestrial surface gravity).  We have not yet investigated the 
variety of chemical and thermal results of such irradiation, but the 
deposition will likely result in a ``secondary ionosphere" at 
altitudes around 5-20 km, and substantial photochemistry (e.g. NO$_x$ 
production and alteration of the ozone abundance as described by Rudermann 1974 
and Crutzen and Bruhl 1996, or other photodissociation-induced 
alterations).  
We note that a GRB that is near enough to provide a biologically
interesting fluence at the top of the atmosphere of $10^5$ erg cm$^{-2}$
(see below) provides the same energy as 1 - 10 year's worth of 
present-day ambient cosmic
rays (Ruderman 1974; Ellis \& Schramm 1995).  If the secondary electrons
from the \gr\ Compton scattering create an ionization cascade similar
to cosmic ray protons then a GRB at this fluence will produce about
about $3\times10^{14}$ ion pairs cm$^{-2}$ (Ruderman 1974) in the secondary
ionosphere and of order $10^7$ NO molecules cm$^{-2}$ via 
N$_2$ dissociation (Ellis \& Schramm 1995).
We postpone an examination of these processes to a later 
paper and concentrate on the direct biological effects here. 

\subsection{Critical Fluence for Biological Effect}

In order to understand the possible biological importance of
GRBs, the critical  $\gamma$-ray and hard X-ray (together
referred to as ``ionizing radiation" below) as well as UV 
fluences that are typically required for significant DNA
alterations must be estimated.  It is generally agreed that
damage at all large-scale levels (e.g.~organisms, organs)
traces to cell damage, and that damage to DNA, in the form of
single and double strand breaks, base damage, combinations of
complex breaks, and cross-linking within DNA or with protein, is
apparently implicated in most cellular effects, including
mutation, chromosome aberrations, and cell killing and
transformation (see papers in Fielden and O'Neill 1991; for
modern textbook accounts see Friedberg, Walker, \& Siede 1995;
Alpen 1998).  For the energy range where photoabsorption sets in 
(50--100 keV), the main
effect is ionization, which weakens or breaks valence bonds;
also, any unpaired electrons left in covalent bonding will be
very reactive and can cause cross-bonding of molecules,
synthesis of new molecules, or polymerization.  At 
energies above about 0.5 MeV, Compton scattering dominates, and then, above
about 1 MeV, pair production can occur.  Additional aspects of
the problem include, for example, the effect of oxygen in
increasing the photosensitivity of biological material (the
``oxygen effect'').  These ``dose-modifying'' processes should
not affect our order-of-magnitude estimates (for a review of
radiosensitization, see van der Schans 1991).

For ionizing radiation, biological damage is almost always quantified 
in terms of the
dose of radiation absorbed, usually in units of rads (1 rad =
100 erg g$^{-1}$ absorbed); contemporary literature uses the Gray, 1
Gray = 100 rad.  The photon fluence (in erg cm$^{-2}$) required
to produce 1 Gray of damage depends on the energy stopping power or
linear energy deposition (``LED,'' erg cm$^{-1}$ here) and the density
of the material of interest, but mostly on the energy-dependence
of the absorption cross section.  For example, a dose
of 1 Gray is equivalent to roughly $5\times10^5$ erg cm$^{-2}$ for photon
energies between 0.1 and 100 MeV in water (Andrews 1974; Anderson 1984;
Turner 1996), but the conversion
factor decreases rapidly for smaller photon energies.
We realize that the biological effects cannot be
quantitively parameterized simply in terms of the absorbed dose
(spatial distribution of damage, temporal phase of cell
activity, and other factors are also significant), but the
``quality factor'' and ``relative biological efficiency'' that
are introduced to account for these variations are usually of
order one to a few, although larger variations do occur.  It
should also be noted that total absorbed dose (or fluence) is
not the only important factor in biological radiation damage:
the dose rate (which can be converted to a flux, as done above for
absorbed doses and fluences) may be of crucial significance.
For example, damage fluxes may be bounded on the low side partly
because of the existence of cell repair mechanisms that only
have time to operate when the flux is small enough.

An interesting and useful
result for our perspective is that a large variety of eukaryotic cellular
and whole organism damage by ionizing radiation occurs for a fairly narrow range of
absorbed \gr\ and X-ray doses.  A number of studies of DNA
single- and double-break damage, chromosomal aberrations,
cross-linking, and other types of damage induced by exposure of
mammalian cells and human lymphocyte cells to 5--100 keV X-rays
and MeV \grs\ all suggest an effective dose of 1 -- 10 Gray
for significant damage or lethality (see papers by Iliakis et al.
and by Radford, Frankenberg, Sasaki, \& Edwards in Fielden and O'Neill 1991;
also Bird et al.~1980; Geard 1982; Wilson et al.~1993), which
corresponds to a fluence of $5 \times 10^5$ to $5 \times10^6$
erg cm$^{-2}$ for \grs. 
Surprisingly, the summary of the Life Span Study of Hiroshima and
Nagasaki survivors given by Turner (1995, ch.13) indicates a doubling
dose for phenotypical mutations also around 1 Gray.
The level of ``damage'' required
for significance in biological evolution is highly uncertain,
and may be much smaller.  For example, the dose of ionizing
radiation required to induce mutations at a rate equal to the
spontaneous mutation rate, the ``mutation doubling dose," is
typically 0.1 to 0.3 times the mean lethal dose.
 For this reason we think that a
conservative estimate of the critical fluence is the lower
limit of \gr\ damage fluences quoted above, $\rm{F_{cr,\gamma} = 5
\times 10^5~erg~cm^{-2}}$.  This number was basically derived
for mammalian cells, but in the subsequent discussion we will
refer to this as the critical fluence for eukaryotic cells.  

        The lethal dose for prokaryotes is less clear.  Gamma-ray and
X-ray sterilization devices commonly use 20,000 Gray, but this enormous
value mostly reflects the two facts that: 1) The lethal doses for
ionizing irradiation of bacterial spores (e.g. Russell 1982, ch.4)
and viruses (Rohwer 1984) are extremely large, approaching that of
the extremely radiation-resistant non-sporing bacterium {\it
Deinoccocus radiodurans} ({\it D. rad}); and 2) the irradiation is intended to
reduce the population by a factor of $10^6$, not 1/e as in the D37
dose. Archaean hyperthermophiles (Kopolov et al. 1993; DiRuggiero et
al. 1997) are also apparently extremely radiation resistant, although
less so than spores and viruses.  Farkas (1998) notes that, under
specific conditions, a 150 to 700 Gray dose is sufficient for the
control of most non-sporeforming bacteria.  Various studies of
wild-type and numerous mutant forms of {\it Escheria coli} yield doses
for 90 percent reduction in the population (D10 dose) of about 300 
Gray (Raikowski \& Thayer 2000; Tayer \& Boyd 1993; Dion et al. 1994;
Lee et al. 1999).  A study of seven food pathogenic bacteria
irradiated with $\gamma$-rays yielded D10 values ranging from about 300 
Gray down to about 30 Gray (for {\it Vibrio parahaemolyticus}, a
halophilic Gram-negative bacillus).  (Some microbiology texts, e.g.
Black 1999, p. 332, state that many bacteria are killed by doses of
only a millirad ($10^{-5}$ Gray) of ionizing radiation; we can find no
support for this statement in the literature.) Thus it appears that
the critical fluence adopted above is applicable only for eukaryotic
organisms, and should be at least 10 to 100 times larger for bacteria.
Bacterial spores, members of Archaea, and extremely resistant
bacteria like {\it D. rad} require much larger doses.  On the other
hand, even ``typically resistant" prokaryotes have highly complex DNA repair
pathways that may have not been available to the earliest lifeforms 
or to extraterrestrial microorganisms.  These considerations underline
the fact that out estimates must be considered extremely uncertain,
especially for extraterrestrial and early terrestrial organisms.
More positively, primitive eukaryotic branching may have occurred much earlier
than had been supposed before the availability of complete genome
sequences (see the concise review by Graham et al. 2000 and
references therein), so that our estimates of the mean time between
events given below for eukaryotic critical fluences
should still be applicable to many forms of life on Earth (and
elsewhere if there are extraterrestrial equivalents of eukaryotes)
over most of the Earth's history.
We give estimates for both eukaryotes and prokaryotes below.

        Concerning  the critical fluence for UV biological effects, we 
refer the reader to Scalo, Wheeler \& Williams (2001) for a summary. 
As discussed there, the best-studied prokaryotic systems (E. coli and 
B. subtilis) have lethal doses around $10^{4}$ erg cm$^{-2}$, but there 
are significant variations in both directions.  Scalo et al. argue 
that the prevalence of non-lethal but error-prone repair of 
cyclobutane pyrimidine dimers (the primary UV photoproduct) implies 
that the mutation doubling dose (which is our main interest since we 
are not primarily concerned with sterilization events) may be much 
smaller, and they suggest an estimate of only 0.02 erg cm$^{-2}$ for 
this quantity.   Given the uncertainties, and to be conservative, we 
adopt a much larger value only ten times smaller than the lethal 
dose, F$_{cr,UV} = 1\times10^{3}$ erg cm$^{-2}$.  For the atmospheric
attenuation we adopt the calculations of Cockell (1998) for the
absorption and scattering in CO$_2$ and N$_2$ atmospheres that
give about a factor of 3 in the 200 - 300 nm region.  This is 
a lower limit since it ignores non-ozone UV shields such as
aerosols.  Adopting an attenuation factor of 3 gives a critical
flux above the atmosphere of F$_{cr,UV} = 3\times10^{3}$ erg cm$^{-2}$.

We thus estimate that the critical flux to cause biological influence
on a planet with a thin atmosphere, such as Mars with a column
density of about 10 g cm$^{-2}$, is F$_{crit,\gamma} = 5\times10^5$ erg cm$^{-2}$
for eukaryotes and a perhaps a factor of 100 larger for prokaryotes.  For thick
atmospheres like the Earth we assume, for illustration, that 1 percent
of the incident fluence arrives at the ground as biologically-relevant
UV.  For a critical fluence at the ground of 
F$_{crit,UV} = 3\times10^3$ erg cm$^{-2}$, the critical fluence
at the top of the atmosphere is F$_{crit,UV} = 3\times10^5$ erg cm$^{-2}$,
roughly equivalent to the critical fluence for eukaryotes in 
thin atmospheres.
 
\subsection{Results}

The distance at which a GRB can produce the critical \gr\ fluence 
for eukaryotes is about 14 kpc and for prokaryotes about 1.4 kpc. 
The distance to affect the surface of a planet with a thick atmosphere
with UV is about 11 kpc.  GRBs anywhere in the Milky Way
may be significant.  
The average nearest event distance (Eqn. B4) is 
$\rm{\bar{\ell} = 740~pc~t_{Gyr}^{~~-1/2}~pc}$ for the
intermediate evolution case (520 and 1900 pc for the no
evolution and strong evolution cases).   These mean distances
are much smaller than the critical distances 
just derived.  It is thus clear that
biologically significant photon ``jolts'' from GRBs must have
been frequent during the Earth's history.  Note that, as for the
case of chondrule heating, this result depends only on the ``observed"
rate and fluence of the GRBs and hence is independent of collimation.

Using Eqn. (B8) for the mean time between significant events,
\begin{equation}
\rm{T = \frac{4F_{cr}~A}{<L_{peak}>\Delta t_\gamma S_{MW}}} = 7.5~yr~F_{cr} A,
\end{equation}
where $F_{cr}$ is in units of erg cm$^{-2}$, A is the attenuation
factor by which the incident flux is depleted, and the numerical coefficient 
on the RHS is for the intermediate evolution case.  For no and strong 
evolution, the value would be 4.6 and 23 respectively.
.  For UV we are taking
$A_{UV} = 3$ and for \grs\ $A_{\gamma}$ = exp(N/35) where N is the 
column depth in gm cm$^{-2}$.  We find $\rm{T = 2- 4 \times 10^6~yr}$ for the
intermediate evolution case for eukaroytic organisms in thin
atmospheres or UV exposure in thick atmospheres.  
For prokaryotic organisms these times
would all be increased by about two orders of magnitude.  Thus for the
intermediate evolution case {\it at least 1000 biologically significant
Galactic UV irradiations should have occurred
stochastically during the 4.4 Gyr history of life on the Earth.}  
For strong cosmological evolution, which is favorred by some 
recent interpretations (see Hogg 2001), this number of irradiations 
would be decreased by a factor of three.  With the current thin atmosphere
of Mars, eukaryotes on the Martian surface would be exposed to 
lethal bursts at the rate of 400 Gyr$^{-1}$.  While Mars may
have had a thicker atmosphere in the past (see Haberle et al. 1994),  
eukaryotes should have been exposed
to biologically significant bursts of \gr\ irradiation 
many times since it lost most of its atmosphere. 

It is interesting to note that only one hemisphere of the target planet would
have been irradiated at the time of each event, since the event
duration is of order 10 sec (neglecting the extended, but less
powerful afterglow).  Interesting consequences also follow by
considering the protection afforded by various coverings.  For
example, the opacity of water at 100 keV is about 0.02 cm$^2$
g$^{-1}$, so organisms below about 50 m of water would have been
protected.  For surface materials (e.g.~rocks, leaves) the
densities are larger, so the shielding depth
is smaller.  An interesting, but very speculative,
possibility is that the long-term viability of surface-dwelling
organisms might have only been possible because of an unusually
long lull between the stochastic photon irradiations.  At
present there is little known about why the transition of life
from ocean to land occurred, although a common speculation is
that the transition required the development of a significant
ozone layer due to oxygen injection by bacterial photosynthesis
and geological erosion.  Currently the timing of the two events
is too uncertain to test this idea.  The present work
offers an alternative explanation --- a lucky lull in nearby Galactic
activity.

Mutational evolution may be as interesting, if not more interesting, than
intermittent lethality.  Non-catastrophic
hypermutation events may not, however, have any significant evolutionary
effects,   even if they occur much more frequently.  The problem is the
short duration of the burst.  Each X-ray or UV-line jolt will increase the
short-term diversity of all exposed genomes through mutation, but,
integrated over a long time, the number of mutations is insignificant
compared to the integrated number of spontaneous mutations or
exogenous mutations due to terrestrial sources (e.g. soil
radioactivity).  The intermittent exposures could be important if
their duration were long enough to allow the mutations to spread
through an entire population (fixation), but, as we argue
elsewhere (Scalo, Wheeler \& Williams 2001; Scalo, Williams \& Wheeler
2001), the minimum time for fixation is probably a few hundred
generations for bacteria, which corresponds to a week for
the generation time of 1000 s for {\it E. coli}.  GRBs have too small a
duration to allow for mutational fixation, and would therefore only
represent a small fluctuation in the time-integrated genomic
diversity on which natural selection, neutral evolution, or any other
concept for evolution, can operate.
  
The only exception we know to this conclusion is the possibility of 
``adaptive mutations," meaning adaptively-positive mutations that 
occur during the resting state of the cell, i.e. before reproduction 
occurs; in such a case there would be no time constraint.   
Much of the relevant evidence on adaptive mutation can, however, be more 
plausibly interpreted in terms of ``mutator genes" which increase the 
spontaneous (mostly deleterious) mutation rate in response to 
environmental stress; see Forster (1999) for a review and Scalo et al. 
(2001) for further discussion in the context of extraterrestrial events.
      
For these reasons it seems most likely that, if intermittent
GRBs have any significant effect on biological evolution,
it must be through direct lethality or by means of alteration of
atmospheric chemistry.  The latter is capable of severe modification
of niche structure that is believed to be crucial  in, for example,
speciation.  For example, the erosion of a planetary ultraviolet
shield (e.g. ozone) by the propagation of \grs\ through the
atmosphere could lead to significant alterations in the UV flux
incident at the surface.  Although it is not possible at present to
say how long it should take for such departures from equilibrium
photochemistry to return to equilibrium, or to a different equilibrium,
it is surely much longer than the duration of the GRB.
If the time to return to atmospheric chemical equilibrium
exceeds the time required for fixation and adaptation of
mutations in bacterial populations, e.g. weeks, then evolutionary
effects of GRBs could occur indirectly, by means of enhanced
UV mutations allowed by the altered atmospheric chemistry.
Many other scenarios could be imagined, for example those involving
hydrocarbon photochemistry-dominated atmospheres, rather than ozone.
Future calculations of sudden departures of atmospheric chemistry are
required to settle this question.

We note that the biologically-significant fluence adopted above
does not apply to all organisms.  For example, the bacterium
{\it D. rad} and other members of its genus are remarkably resistant to 
extremely large doses of ionizing and UV radiation, because of
exceedingly efficient DNA repair (see Minton 1994, Battista
1997 for reviews).  The critical fluences for \gr\ and UV
damage are about a factor of 10$^2$ larger than adopted here for
``normal'' bacteria.  Using the numbers given earlier, it is then
likely that {\it D. rad} has been exposed to lethal doses from 
GRBs more than a few times during the history of the Earth,
assuming that it is deep branching.  
A common statement in the literature on {\it D. rad}
is that it has never been subjected to natural radiation fluences 
equal to its tolerance, so that its radiation resistance must be a 
by-product of some other repair pathway, e. g. for dessication 
resistance (see Battista 1997).  The expected fluence of GRBs shows 
that this is not true. The possible additional importance 
for {\it D. rad} of UV radiation from supernovae during 
their light curve evolution will be discussed elsewhere.

Finally, we point out that although we have found GRBs to be
of potential biological importance as sources of stochastic
irradiation events, it is by no means clear that they are the
most important.  Other events, such as the ultraviolet burst
associated with SN shock breakout, the ultraviolet radiation
and radioactive decay \grs\ from SN light curves, soft
\gr\ repeaters, flare stars, or massive
spectral type O stars may make important or even dominant
contributions to this Galactic background of stochastic
irradiation events.  We postpone a discussion of these other
sources to a separate publication.

\begin{center}
Acknowledgements
\end{center}

We thank Peter H\"oflich for clarifying important atomic processes 
for us and Peter H\"oflich and David Smith for carrying out 
preliminary Monte-Carlo simulations to 
estimate the net transfer efficiencies.  We also thank
Charles Cockell, Andy Ellington,
and Andy Karam for stimulating
discussions on the biological aspects of this work and Brad Schaefer
and Pawan Kumar for discussions of GRBs.  We appreciate comments
by two referees that caused us to consider the alternative
surface brightness approach for estimating the GRB rates
per unit blue luminosity, to clarify the organization of the
paper, and to give more consideration to the potential
evolutionary effects.  This research was supported by NSF grants AST 981960,
9907582, and 9903582.

\newpage
\begin{center}
\textbf{APPENDICES}
\end{center}
\begin{appendix}
\section{METHOD FOR ESTIMATING GRB RATES BASED ON INTEGRATED
BACKGROUNDS}

Another method to estimate the desired GRB rate (suggested by
an anonymous referee) is, aside from selection effects,
rather independent of either the peak luminosity distribution
or the redshift distribution of the GRBs.  Using only the
assumption that the rate of GRBs is proportional to
the SFR and that the blue light produced by
galaxies is also proportional to the SFR, one can
construct the ratio of the total GRB rate per unit solid angle
of the sky and compare that to the integrated background blue
light from galaxies per until solid angle.

We have examined the distribution of GRB number vs peak flux given
by Fenimore \& Bloom (1993), who give the integral distribution,
and that of Totani (1999), who gives the differential distribution.
To compare below to the analysis based on the results of
Schmidt (1999), we integrate over the BATSE response
of about 1 to 100 ph cm$^{-2}$ s$^{-1}$. The results
are not sensitive to these limits.
With this choice of the limits, we can evaluate the integrated
count rate as:
\begin{equation} 
    \int_{1}^{100}Pf(P)dP = 2000~{\rm ph~cm^{-2}~s^{-1}~yr^{-1}},
\end{equation} 
where $ f(P) = dN/dP$ is the differential distribution of
GRBs with detected peak flux P. Note that the units ph s$^{-1}$
refer to peak count rates and the units $\rm yr^{-1}$ refer to
the rate of events with a given peak flux per year.
For similar limits, integrating Totani or integrating Fenimore
\& Bloom give similar numbers, to within 50 percent.

For an E$^{-2}$ spectrum, the average energy of a photon in the BATSE
band is $<\rm{E}> 
= 1.72\times10^{-7}$ erg ph$^{-1}$ (e.g. Schaefer, Deng \& Band
2001).   We can define the quantity
\begin{equation} 
   Q_{\gamma,\Delta\Omega}
= <E>\frac{4\pi}{\Delta\Omega}\int_{1}^{100}Pf(P)dP
= 8.3\times10^{-5} ~{\rm erg~cm^{-2}~s^{-1}~yr^{-1}~Sr^{-1}}
\left(\frac{\Delta\Omega}{4\pi/3}\right)^{-1},
\end{equation} 
where we have assumed the average sky coverage of BATSE
is about 1/3 of $4\pi$ sterradians.  The quantity $Q_{\gamma,\Delta\Omega}$
is the analogy to the quantity $Q_{\gamma,V}$ defined in \S2 (cf. Eqn. 2).
The quantity $Q_{\gamma,\Delta\Omega} \Delta t_{\gamma}$ is
the power detected in $\gamma$-rays per unit time per
unit area of detector per unit solid angle, where $\Delta t_{\gamma}$ 
is the mean duration of a GRB, $\sim10$ sec (\S2).

We take the galaxy B band luminosity background from Pozzetti et al.
(1998; see also Madau \& Pozzetti 2000) to be:
\begin{equation} 
   J_{B,\Delta\Omega} =  4.6\times10^{-6}~ {\rm erg~cm^{-2}~s^{-1}~Sr^{-1}}.
\end{equation} 
Madau and Pozzetti note that faint objects that are brighter than the
nominal depth of the catalog may be missed due to the
$(1+\rm z)^4$ dimming factor and that there may be a correction
for unresolved galaxies.  This number is thus a lower limit.
In addition, the typical galaxy in the sample has a redshift of
$\sim 0.6$, whereas the typical redshift of a GRB may be $\sim 2$.
A volume correction proportional to (1 + z)$^3$) could thus be
a factor of order 6.  The implication is that while this method
is independent of redshift distribution in principle, it is
subject to the redshift distribution in practice through selection
effects.

We can now form the ratio
\begin{equation} 
    \frac{Q_{\gamma,\Delta\Omega}\Delta t_{\gamma}}{J_{B,\Delta\Omega}}
\end{equation} 
which is the energy received in $\gamma$-rays per unit energy
detected in integrated blue light from background galaxies.
If we multiply by the surface brightness of blue light for
a typical galaxy, in our case that for the Milky
Way, $\Sigma_{\rm{L,B}}$ = $8\times10^{34}$
erg s$^{-1}$ pc$^{-2}$, we can form the quantity
\begin{equation} 
F_{\Delta\Omega} =  \frac{Q_{\gamma,\Delta\Omega}\Delta t_{\gamma}}
{J_{B,\Delta\Omega}}
 \Sigma_{\rm{L,B}} = 4.6\times10^{29}\left(~\frac{\Delta
t_{\gamma}}{10~s}\right)
\left(\frac{\Delta \Omega}{4\pi/3}\right)
{\rm [~erg~of~\gamma-rays~s^{-1}]~pc^{-2}}.
\end{equation} 
The quantity F$_{\Delta\Omega}$ is the rate of production of energy
per unit area by GRBs in a galaxy and is equivalent to $F_V$.
>From Eqn. (\ref{FV}) in \S 2,
we have F$_V$ = 1.6$\times10^{29}$
${\rm [~erg~of~\gamma-rays~s^{-1}]~pc^{-2}}$, so the two
methods roughly agree, neglecting potential uncertainties in both.

Whereas $Q_{\gamma,V}$ gives the product  $\rm{r_0 <L_{peak}>}$,
the analysis in terms of background surface densities
directly gives the product $\rm{S_\gamma <L_{peak}>}$, itself a useful
quantity to determine the mean time between GRBs, as shown in Appendix B
(see also
\S6, Eqn. 10).   This determination of background fluxes cannot provide the
quantities
$\rm{S_\gamma}$, the GRB rate per unit galaxy area, or ${<\rm L_{peak}>}$
separately.  Although this method does provide an estimate of
the mean time between GRBs that is relevant for biological applications
(\S 6), even in this sort of application,
one separately needs $\rm{S_\gamma}$ and $\rm{E_\gamma}$
= $\rm{<L_{peak}>}$$\rm{ \Delta t_{\gamma}}$ to compute the
quantities such as the mean nearest distance and the critical
distance that will give a threshold fluence.  The background
flux method cannot be used to compare GRB rates to those of
supernovae (\S 3) where $\rm{S_\gamma}$ is needed or HI holes (\S 4) since
again one
needs  direct information of $\rm{E_\gamma}$.  Quantities that come into
the estimation of the heating of chondrules require the
independent knowledge of $\rm{S_\gamma}$ alone, especially
to estimate the probability of more than one GRB heating event.
Thus although the method based on integrated background
fluxes provides an independent check on some aspects of
the problem it is subject to selection effects and, in addition,
our applications require the information provided
in the approach of Schmidt that cannot be obtained by this method.

\newpage

\section{A POISSON MODEL FOR FLUENCE PROBABILITIES}

For some of the applications of interest it is necessary to
estimate the probability that a GRB occurred within a distance
$\ell_{\rm{cr}}$, such that the fluence received at a given point
exceeds a critical value F$_{\rm{cr}}$ = E$_{\rm{s}}$ /
4$\pi\ell_{\rm{cr}}^{~~2}$, where E$_{\rm{s}}$ is the total
energy emitted by the burst.  Although the GRB events in a given
galaxy are likely to be clustered, in order to make such an
estimate we assume that the GRB events are randomly distributed
in space and time and can be described by a Poisson spatial
process.  If we observe the process over time, and mark the
distance of the nearest event, this nearest distance decreases
with time, as the number of ``markers'' increases within a given
volume or area.  We are therefore interested in the probability
distribution for the distance of the nearest event as a
function of time.  It will be seen that the accumulated nearest
event distances are large compared to a galactic scale height,
because of the relatively small rate of GRBs.  Thus we consider
a two-dimensional Poisson process.  (By contrast, supernova
events are frequent enough that the appropriate distribution of
markers would be three-dimensional.)  We also make our
estimates using the mean GRB energies and total rates, rather
than including the effect of the GRB luminosity function on the
calculation, since we are interested in order of magnitude
results at this point.

Let S$_\gamma$ be the rate of GRB events per unit area and A the area of
interest.  The number of accumulated events per unit area after
time t is $\nu$ = ${\rm S_{\gamma} t}$.  Then the probability that k events
have
occurred in an area A during time t is
\begin{equation}
\rm{P (k) = (\nu A)^k~exp(-\nu A)/k!}
\end{equation}
Let $\ell$ be the distance to the nearest marker.  The probability
that a circle of area A contains zero markers is
P(0)=exp($-\nu$A).  This is equivalent to the probability
that the first (i.e. nearest) marker occurs at a distance
greater than that corresponding to area A.  The value of
P(0) is thus the cumulative probability distribution
of nearest distances corresponding to area A,
$\Phi(> A)$.  By definition,
\begin{equation}
\Phi(> A) = \int^{\infty}_{A}\phi(A)dA,
\end{equation}
\noindent where $\phi(A)$ is the differential distribution or
probability distribution function (pdf) of nearest
distances corresponding to A.  The function $\phi(A)$ is
obtained by differentiating $\Phi(> A)$ as
$\phi(A) = \rm{\nu exp}(-\nu A)$.  Transforming this pdf to the
pdf of the nearest distance, p($\ell$) = $\phi$(A($\ell$)),
where A = $\pi \ell^2$, gives
\begin{equation}
\rm{p (\ell) = 2 \pi \nu \ell~exp (-\pi \nu \ell^2)}.
\end{equation}
\noindent In statistics texts this is usually given as the
distribution of nearest neighbors, but it is clear that it is
equivalently the probability distribution of nearest distances
from any point.  By integration, the mean nearest distance is
\begin{equation}
\rm{\bar{\ell} = (S_\gamma t)^{-1/2}/2}.
\end{equation}
\noindent For the intermediate redshift evolution parameters (\S2),
we find that $\bar{\ell} = 740~\rm{t_{Gyr}^{~~-1/2}~pc}$, where
t$_{\rm{Gyr}}$ is time in units of 10$^9$ yr.  For the no
evolution and strong evolution cases, the numerical coefficient
is 520 pc and 1900 pc, respectively.

The probability that a GRB has occurred at a distance less than
$\ell_{\rm{cr}}$ (corresponding to the critical received fluence
F$_{\rm{cr}}$) in time t is obtained by integrating p($\ell$) from 0
to $\ell_{\rm{cr}}$.  The result is
\begin{equation}
\rm{P(\ell<\ell_{cr})=1 - exp[-(\pi/4)(\ell_{cr}/\bar{\ell})^2]}.
\end{equation}
The argument of the exponential is
\begin{equation}
\rm{(\pi/4)(\ell_{cr}/\bar{\ell})^2=\frac{<L_{peak}> \Delta t_\gamma
~S_\gamma t}{4F_{cr}}}
\end{equation}
\noindent where the mean energy release per event has been
written as the mean peak luminosity $\rm <L_{peak}>$ times some average
duration $\Delta$t$_\gamma$ ($\approx$10 sec, see \S 2), and
\begin{equation}
\rm{\ell_{cr} = (<L_{peak}> \Delta t_\gamma / 4 \pi F_{cr})^{1/2}},
\end{equation}
\noindent is the critical distance for an event that results in
a a fluence F$_{\rm{cr}}$.  Equating the argument of the
exponential in Eqn. (5) to unity gives the average time between events
that produce a received fluence F$_{\rm{cr}}$ as
\begin{equation}
\label{meantime}
\rm{T =\frac{ 4F_{cr}}{<L_{peak}> \Delta t_\gamma S_\gamma}}.
\end{equation}
To indicate the dependence on the cosmic SFR history,
we take for illustration a fiducial critical fluence of 10$^9$
erg cm$^{-2}$, giving $\ell_{\rm{cr}}$ = 300
F$_{\rm{cr,9}}^{~~-1/2}$ pc.  This result is changed by less
than fifty percent if we consider the no evolution and strong
evolution cases.  The mean time between significant
(F$>$F$_{\rm{cr}}$) events in the Galaxy is, from Eqn. (\ref{meantime}),
T = 5.4 $\times~10^9
~\rm{F_{cr,9}}$ yr for the intermediate evolution case.  For no
evolution and strong evolution, the numerical coefficient
becomes $3.8 \times 10^9$ and $3.5 \times 10^{10}$,
respectively.  The strong (no) evolution case gives larger
(smaller) interevent times because the local GRB rate is
smaller (larger) in that case.

The above derivation approximates the full LF of the peak
luminosities of GRBs by
the {\it mean} peak luminosity.  More realistically, the above
probabilities would represent conditional probabilities for a
specified luminosity, which would then have to be integrated
over the LF to find the probability for a given critical
fluence.  For example, Schmidt's (1999) adopted LFs would give
a larger number of events below the mean peak
luminosity, somewhat reducing the estimated probabilities and
increasing the derived average time.  On the other hand, for
Kumar \& Piran's (2000) stochastic model, the effective LF is
much more symmetrical about the mean, with a significant
fraction of events at larger and smaller luminosities than the
mean, suggesting that our estimates based on the mean
luminosity would be essentially unchanged.

\end{appendix}

\newpage

\begin{deluxetable}{lccccc}

\tablecolumns{6}

\tablecaption{Properties of Gamma-Ray Bursts}

\tablehead
{\colhead{Case}
&\colhead{\scriptsize$\rm{r_0}$}
&\colhead{\scriptsize$\rm{r_B}$}
&\colhead{\scriptsize$\rm{<L_{peak}>^{~~~(a)}}$}
&\colhead{\scriptsize$\rm{S_{MW}^{~~~(b)}}$}
&\colhead{\scriptsize$\rm{S_{MW}<L_{peak}>\Delta t}$}\\
\colhead{}
&\colhead{\scriptsize$\rm{Gpc^{-3}~yr^{-1}}$}
&\colhead{\scriptsize$\rm{10^{-17}~l^{-1}~yr^{-1}}$}
&\colhead{\scriptsize$\rm{10^{51}~erg~s^{-1}}$}
&\colhead{\scriptsize$\rm{10^{-16}~pc^{-2}~yr^{-1}}$}
&\colhead{\scriptsize$\rm{10^{36}~erg~pc^{-2}~yr^{-1}}$}}

\startdata
No evol.             &6.6~${\rm h_{70}^3}$   &4.4~${\rm h_{70}^2}$
&0.9~${\rm h_{70}^{-2}}$    &9.3~${\rm h_{70}^{2}}$    &8.4\\
Intermediate evol.   &3.6~${\rm h_{70}^3}$   &2.4~${\rm h_{70}^2}$
&1.1~${\rm h_{70}^{-2}}$    &4.6~${\rm h_{70}^{2}}$    &5.1\\
Strong evol.         &0.5~${\rm h_{70}^3}$   &0.3~${\rm h_{70}^2}$
&2.5~${\rm h_{70}^{-2}}$    &0.7~${\rm h_{70}^{2}}$    &1.7\\
\enddata

\tablenotetext{a}{All energies and
luminosities refer to the 10--1000 keV range.}

\tablenotetext{b}{Rate
scaled to the Milky Way using
$\rm{S_{MW}<L_{peak}>\Delta t
= Q \Sigma_{L,B,} \Delta t/J_{gal,B,V}}$.  See text.}

\end{deluxetable}

\end{document}